\documentclass{article}

\usepackage{microtype}
\usepackage{graphicx}
\usepackage{subfigure}
\usepackage{booktabs} 
\usepackage{makecell}
\usepackage{tikz}
\usepackage{amsmath}
\usepackage{siunitx}
\usetikzlibrary{positioning,calc,arrows.meta,matrix,fit}

\usepackage{hyperref}

\usepackage[accepted]{icml2025}

\usepackage{amsmath}
\usepackage{amssymb}
\usepackage{mathtools}
\usepackage{amsthm}
\usepackage{makecell}

\usepackage[capitalize,noabbrev]{cleveref}

\theoremstyle{plain}

\theoremstyle{definition}

\theoremstyle{remark}

\usepackage[textsize=tiny]{todonotes}

\icmltitlerunning{WhisperKit}

\begin{document}

\twocolumn[
\icmltitle{WhisperKit: On-device Real-time ASR \\ with Billion-Scale Transformers}



\icmlsetsymbol{equal}{*}

\begin{icmlauthorlist}
\icmlauthor{Berkin Durmus}{sch,comp}
\icmlauthor{Arda Okan}{comp}
\icmlauthor{Eduardo Pacheco}{comp}
\icmlauthor{Zach Nagengast}{comp}
\icmlauthor{Atila Orhon}{comp}
\end{icmlauthorlist}

\icmlaffiliation{comp}{Argmax Inc., Palo Alto, USA}
\icmlaffiliation{sch}{University of California, Los Angeles, USA}

\icmlcorrespondingauthor{Atila Orhon}{a@argmaxinc.com}

\icmlkeywords{Machine Learning, ICML}

\vskip 0.3in
]



\printAffiliationsAndNotice{}  

\begin{abstract}
Real-time Automatic Speech Recognition (ASR) is a fundamental building block for many commercial applications of ML, including live captioning, dictation, meeting transcriptions, and medical scribes. Accuracy and latency are the most important factors when companies select a system to deploy. We present WhisperKit, an optimized on-device inference system for real-time ASR that significantly outperforms leading cloud-based systems. We benchmark against server-side systems that deploy a diverse set of models, including a frontier model (OpenAI \texttt{gpt-4o-transcribe}), a proprietary model (Deepgram \texttt{nova-3}), and an open-source model (Fireworks \texttt{large-v3-turbo}).Our results show that WhisperKit matches the lowest latency at 0.46s while achieving the highest accuracy 2.2\% WER. The optimizations behind the WhisperKit system are described in detail in this paper.
\end{abstract}

\section{Introduction}

Frontier models have been scaled to trillions of parameters to serve virtually all applications across all modalities with a single model, narrowing down deployment options to the cloud due to unprecedented memory requirements. At the same time, task-focused models, whether distilled or trained from scratch, match or exceed the accuracy of frontier models at a fraction of the inference cost for most applications, making on-device deployment of best-in-class models scalable and economical, even for real-time streaming inference. Given this dichotomy, frontier models have become the choice for fast prototyping and time-to-market with a cloud-based inference API while task-focused models with an on-device inference API become the steady-state solution as the industry matures.

For Automatic Speech Recognition (ASR), Whisper Large v3 Turbo \cite{whisper}, a 1-billion parameter Encoder-Decoder Transformer model, matches or exceeds many frontier models, such as \texttt{gpt-4o-transcribe} \cite{gpt-4o-transcribe}, in ASR accuracy while being sufficiently compact for on-device deployment.  WhisperKit is an optimized on-device inference system designed to deploy Whisper models for real-time streaming transcription on Apple devices. \footnote{\noindent\texttt{Code available at:\url{https://github.com/argmaxinc/WhisperKit}}}.

In specific, WhisperKit makes the following contributions:

\begin{itemize}
    \item We modified Whisper's system architecture such that the \textit{Audio Encoder} model natively supports streaming inference and the \textit{Text Decoder} model yields accurate output text streams even when running on partial audio.
    \item We reimplemented Whisper for native acceleration on the \textit{Apple Neural Engine} (ANE) to achieve near-peak hardware utilization while retaining the required energy efficiency for on-device deployment.
    \item We compress Whisper with a new technique that retains the \textit{Word Error Rate} (WER) within 1\% of the original uncompressed model while reducing the model file size from 1.6 GB to 0.6 GB.
\end{itemize}

\section{Method}
\subsection{Architecture}

Real-time streaming transcription is a challenging ASR task with major commercial applications such as live captioning, dictation, meeting transcriptions, medical scribes, etc. The challenges are twofold: Achieving high accuracy with partial audio context due to the streaming nature of the input while maintaining low latency in real-time processing, where delays compound over time. The latency challenge is exacerbated for model architectures such as Whisper that do not natively support streaming inference. Whisper consists of an \textit{Audio Encoder} and a \textit{Text Decoder}. Both components introduce distinct challenges.

\subsubsection{Audio Encoder}
Whisper Audio Encoder can only process 30-second audio chunks. The naive streaming implementation with a 1-second target latency involves zero-padding the audio input to 30 seconds and running an audio encoder forward pass at most every 1 second as the audio buffer gets updated. For 1 minute of streaming inference, this implementation leads to at least 60 audio encoder forward passes, the equivalent of up to 30 minutes of processed audio, compared to offline inference which requires only a single forward pass.

Moonshine \cite{moonshine} is a rearchitected version of Whisper that removes the need to pad the input to 30-second buffers. Although Moonshine significantly improves the streaming efficiency, it suffers accuracy loss for long-form transcription. Furthermore, the published models were not scaled beyond the \texttt{tiny} and \texttt{base} variants leading to an accuracy gap with larger state-of-the-art models.

\cite{localagreementandunidirectionalencoder} fine-tunes their audio encoder Transformer model with a lower-triangular attention mask such that its self-attention layers become causal. This technique removes the "look-ahead" conditioning caused by the original bidirectional self-attention and enables key-value (KV) caching for incremental encoding of input audio streams.

\begin{figure}[ht]
  \centering
  \subfigure[original]{
    \begin{tikzpicture}[scale=0.2]
      \draw[draw=none] (0,0) grid (6,6);
      
      \draw[thin] (0,0) rectangle (6,6);
    \end{tikzpicture}
    
    \label{fig:original}
  }
  \subfigure[c250]{
    \begin{tikzpicture}[scale=0.2]
      \draw[draw=none] (0,0) grid (6,6);
      
      \fill[black] (5,1) rectangle (6,2);
      \fill[black] (4,2) rectangle (6,3);
      \fill[black] (3,3) rectangle (6,4);
      \fill[black] (2,4) rectangle (6,5);
      \fill[black] (1,5) rectangle (6,6);
      
      \draw[thin] (0,0) rectangle (6,6);
    \end{tikzpicture}
    
    \label{fig:c250}
  }
  \subfigure[d750]{
    \begin{tikzpicture}[scale=0.2]
      \draw[draw=none] (0,0) grid (6,6);
      
      \fill[black] (3,3) rectangle (6,6);
      \fill[black] (0,0) rectangle (3,3);
      
      \draw[thin] (0,0) rectangle (6,6);
    \end{tikzpicture}
    \label{fig:d750}
  }
  \subfigure[d500]{
    \begin{tikzpicture}[scale=0.2]
      \draw[draw=none] (0,0) grid (6,6);
      
      \fill[black] (0,0) rectangle (6,6);
      \fill[white] (0,4) rectangle (2,6);
      \fill[white] (2,2) rectangle (4,4);
      \fill[white] (4,0) rectangle (6,2);
      \draw[thin] (0,0) rectangle (6,6);
    \end{tikzpicture}
    \label{fig:d500}
  }
  \subfigure[d250]{
    \begin{tikzpicture}[scale=0.2]
      \draw[draw=none] (0,0) grid (6,6);
      
      \fill[black] (0,0) rectangle (6,6);
      \fill[white] (0,5) rectangle (1,6);
      \fill[white] (1,4) rectangle (2,5);
      \fill[white] (2,3) rectangle (3,4);
      \fill[white] (3,2) rectangle (4,3);
      \fill[white] (4,1) rectangle (5,2);
      \fill[white] (5,0) rectangle (6,1);
      
      \draw[thin] (0,0) rectangle (6,6);
    \end{tikzpicture}
    \label{fig:d250}
  }
  \label{fig:causalify-masks}
  \caption{
    \textbf{Block-diagonal (d*) and block-causal (c*) masks for Whisper Audio Encoder 1500x1500 (30 seconds) Self-attention Matrix.}
  }

\end{figure}
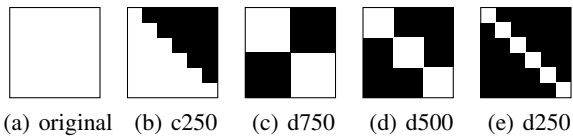

We apply self-distillation to Whisper Large V3 Turbo using the Common Voice 17 \cite{commonvoice:2020} training split following the core ideas from \cite{localagreementandunidirectionalencoder}. In our experiments, we found that the lower-triangular causal attention mask is too strict and led to an accuracy drop. Instead, we relax the attention mask to be block-diagonal where the blocks represent 15-second audio chunks, depicted as \texttt{d750} in Figure \ref{fig:d750}.

After self-distillation of the audio encoder, the text decoder still needs to attend over 30-second audio buffers. However, streaming inference severely underutilizes this buffer length and most of the input buffer needs to be zero-padded, leading to unnecessary compute and latency. Block-diagonal masking enables \textit{silence caching} where the audio encoder's output for an entire 15-second zero-padded block of audio can be computed at compile time and reused instead of running inference on zero-padded inputs.

\begin{table}[ht]
  \begin{center}
    {\scalebox{0.7}{
    \begin{tabular}{lcccccc}
      \toprule
      Mask  & TFLOPs & \makecell{Audio Encoder \\ Latency (ms)} & \makecell{librispeech-test.clean \\ (WER)} & \makecell{earnings22\footnote{We use a random 10\% subset of the earnings22 \cite{delrio2022earnings22} test split. Published dataset link will be added as a footnote upon pulication.} \\ (WER)} \\
      \midrule
      original & 2.27 & 612 & 1.93 & 11.55 \\
      \hline
      c250  & 2.34 & 117 $*$ 6 = 702 & 2.32 (+0.39) & 12.89 (+1.34) \\
      d750  & 1.04 & 218 & \textbf{2.25 (+0.32)} & \textbf{12.85 (+1.30)} \\
      d500  & 0.68 & 125 & 3.39 (+1.46)& 13.86 (+2.31)\\
      d250 & \textbf{0.33} & \textbf{50} & 25.58 (+23.65) & 39.75 (+28.20) \\
      \bottomrule
      \end{tabular}}
    }
    \caption{\textbf{Latency and WER benchmarks for Whisper Audio Encoders self-distilled with various attention masks.} Latency is measured on a MacBook Pro with M3 Max chip on the Neural Engine. \texttt{c250} latency is multipled by 6 because block-causal masking does not allow for \textit{silence caching} and requires 6 forward passes of 5-second audio segments to fill the 30-second audio buffer.
    }
    
  \label{causal-wer}
  \end{center}
  \vskip -0.2in
  \end{table}

\textbf{English Accuracy.} Table \ref{causal-wer} shows the impact of self-distillation using these self-attention masks on the audio encoder latency and WER for short-form and long-form and multilingual transcription. \texttt{d750} retains WER within 1\% of the original model while reducing latency by 65\% (602 ms \textrightarrow 218 ms). Shorter blocks such as \texttt{d250} and \texttt{d500} yield an undesirable trade-off between latency and accuracy.

On the other hand, block-causal masking, depicted as \texttt{c250} in Figure \ref{fig:c250}, retains the same accuracy as \texttt{d750} for 3x shorter audio blocks. However, unlike block-diagonal, block-causal conditioning requires key-value caching for previous blocks in the same 30-second window and does not allow \textit{silence caching} as future blocks within the same 30-second window depend on earlier blocks that change as input audio streams.

  \begin{figure}[ht]
    \centering
    \includegraphics[width=0.45\textwidth]{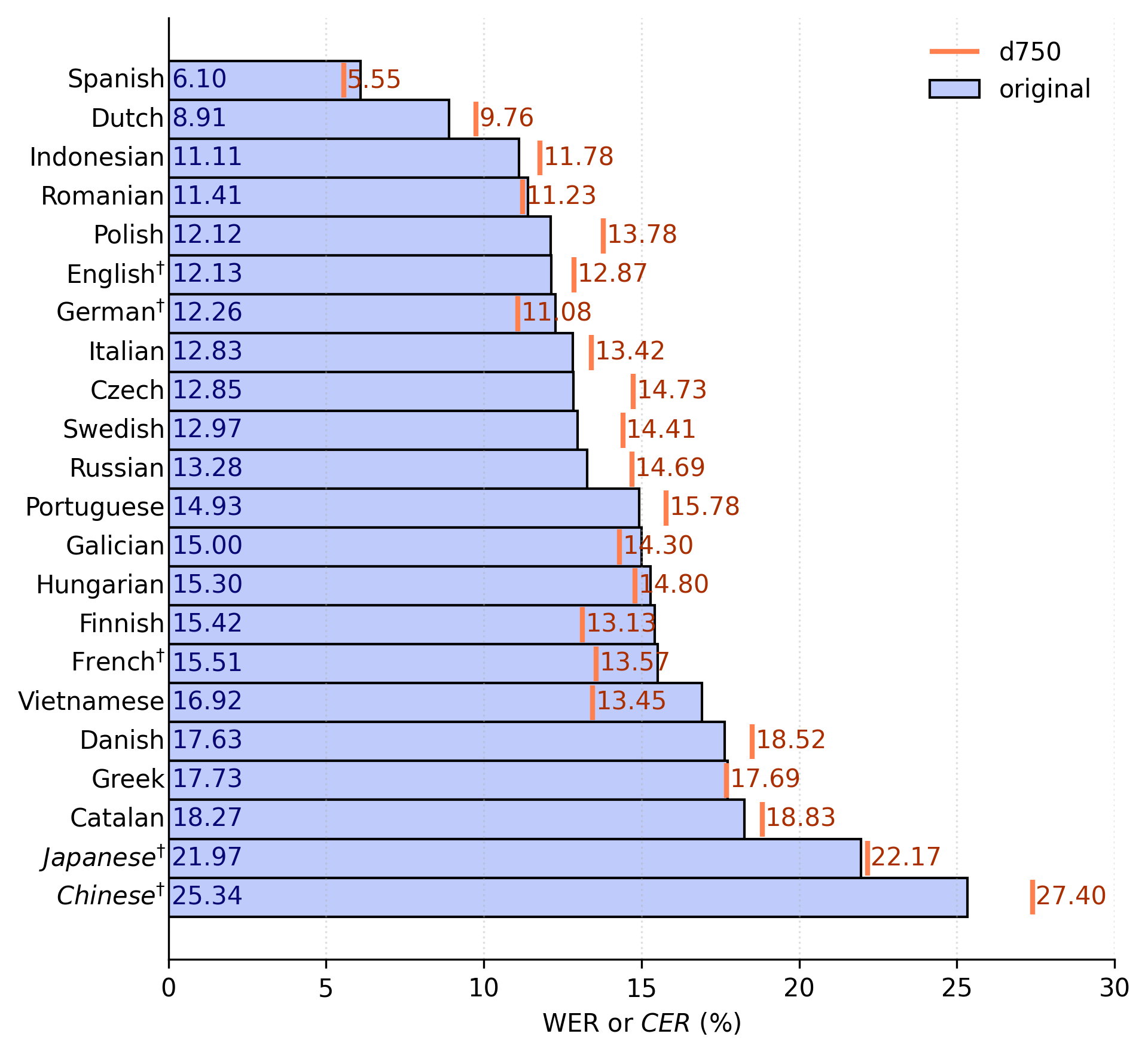}
    \caption{\textbf{Multilingual WER (Word Error Rate) and CER (Character Error Rate) before (original) and after (\texttt{d750}) self-distillation of Whisper Large v3 Turbo.} 
    Metrics are reported for the 20 languages with the lowest WER/CER in the original model, plus Japanese and Chinese, on a subset of the Common Voice 17 test set. \footnote{This random subset of the Common Voice 17 test split will be made available upon publication.}. Languages marked with $\dagger$ were used during self-distillation.
    }
    \label{fig:d750-cv17-wer-bar-chart}
     \vskip -0.2in
  \end{figure}

  \textbf{Multilingual Accuracy.} Figure \ref{fig:d750-cv17-wer-bar-chart} shows the WER/CER of the \texttt{d750} variant on a subset of the Common Voice 17 \cite{commonvoice:2020} test split. Whisper Large v3 Turbo (original) was initially trained to recognize and transcribe in 100 languages. However, the model does not demonstrate useful quality on many of these languages. In order to focus on a useful subset, we focus our benchmarks on the 20 languages where the original model achieves the lowest WER/CER, plus Japanese and Chinese.

  The \texttt{d750} variant was obtained by fine-tuning the original model while applying the self-attention mask from Figure \ref{fig:d750} on a subset of Common Voice 17 training split covering the following 5 languages: English, German, Japanese, Chinese, and French. We hypothesized that the fine-tuned model would achieve even lower WER than the original model for the languages where it was fine-tuned because this strategy allows the model to forget the other languages and recycle its limited parameter count towards fewer languages.
  
  Figure \ref{fig:d750-cv17-wer-bar-chart} shows that the model either retains WER within 1\% (English and Japanese) or improves WER (French and German) when compared to the original model for languages where it was fine-tuned. The exception was Chinese which regressed by 2\% WER, presumably due to small training dataset size.
  
  These results demonstrate that fine-tuning Whisper Large v3 Turbo using this strategy is a scalable approach to improving multilingual accuracy while optimizing inference efficiency.

\subsubsection{Text Decoder}
Unlike the Audio Encoder, the Whisper Text Decoder is capable of streaming individual output tokens while attending over a fixed input audio buffer. However, the naive streaming implementation for the text decoder is also problematic. The output transcript buffer is filled with temporary text tokens, and the buffer is repeatedly flushed and refilled after each audio encoder forward pass until the text decoder can confirm a full transcript buffer by predicting the \texttt{<endoftranscript>} special token. Until this event occurs, all text token predictions are subject to change and the audio cursor can not be moved forward, leading to compounding latency. We call these temporary results \textit{hypothesis text}.

\cite{localagreementandunidirectionalencoder} proposed the \texttt{LocalAgreement} streaming policy that frequently confirms the hypothesis text by searching for the longest common prefix across two consecutive \textit{hypothesis text} buffers and moves the audio cursor to the end of the last token in this common prefix. WhisperKit and concurrent work \cite{whisper-streaming} applied the \texttt{LocalAgreement} policy to Whisper, leading to dual output text streams:

\begin{itemize}
    \item \textit{Confirmed text} stream can be leveraged in the user experience to build trust in stable and accurate results.
    \item \textit{Hypothesis text} preserves low latency and responsiveness with occasional corrections as more audio context becomes available.
\end{itemize}

\textit{Hypothesis text} helps real-time transcription systems achieve sub-second latency while retaining the flexibility for retroactive corrections to match offline transcription accuracy. However, if the number of corrections is high, it may also hurt the user experience and reduce trust in the accuracy of these systems.

\begin{figure}[ht]
  \vskip 0.1in
  \begin{center}
  \includegraphics[width=\linewidth]{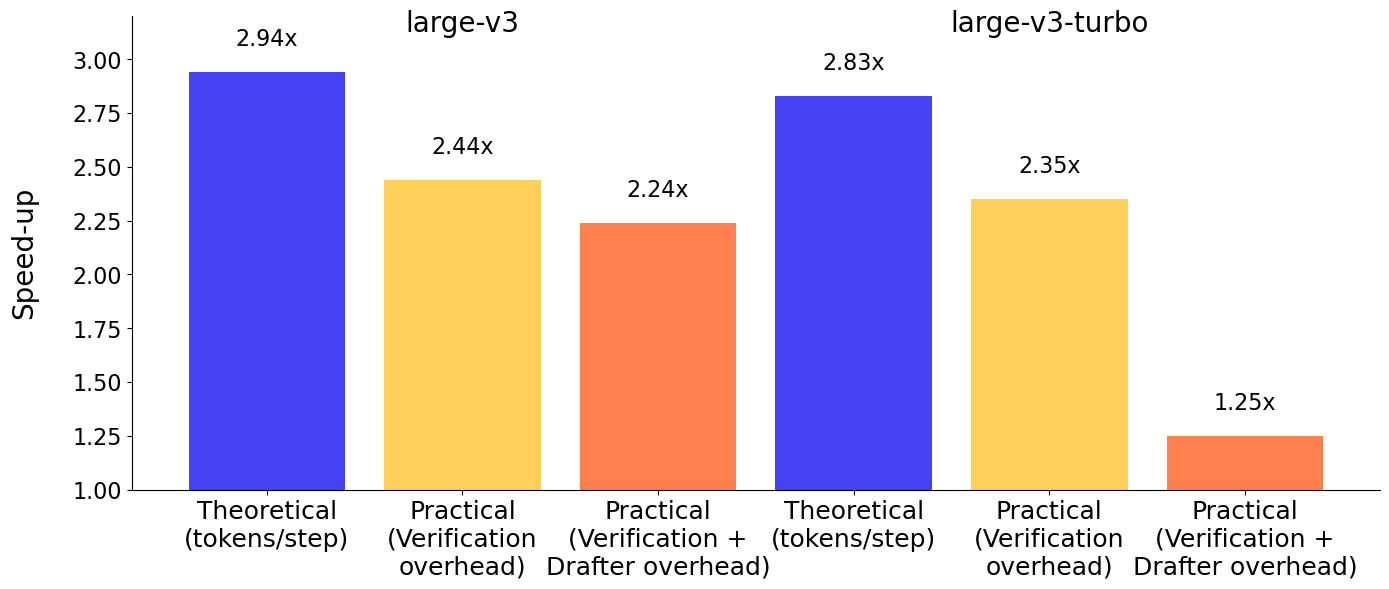}
  \caption{\textbf{Speedup factors for Whisper Text Decoder with Speculative Decoding}. Theoretical speedup (tokens/step) is the token acceptance rate of the verification step and signifies the speedup for an idealized hardware with infinite compute and limited memory bandwidth. The practical slowdowns are due to the \textit{verification overhead} (processing multiple tokens in parallel) and \textit{drafter overhead} (forward pass latency of the drafter model).}
  \label{specdec}
  \end{center}
  \vskip -0.12in
  \end{figure}

\textbf{Speculative Decoding}. \cite{redrafter} introduced \textit{Recurrent Drafter}, a lossless speculative decoding algorithm that uses a lightweight RNN drafter to propose multiple future tokens conditioned on the original model's hidden states, which are then verified in parallel by the original model.

We trained a draft model that achieves a 2.94x and 2.83x theoretical speedup (accepted tokens per step) for Whisper Large v3 and Large v3 Turbo respectively. This theoretical speedup is mostly retained even after accounting for the \textit{verification overhead} (2.94x \textrightarrow 2.41x and 2.83x \textrightarrow 2.35x). The verification overhead is the slowdown of running the text decoder on 16 tokens (beam width multiplied by beam length, $4\times4=16$) in parallel instead of 1 token per forward pass. 

\cite{redrafter} observed that Apple Silicon GPUs achieved significantly lower practical speedups than Nvidia GPUs with identical drafter models due to increased parallel verification overhead which pushes the workload to be compute-bound, especially for large models (7-billion parameters or higher). We show that WhisperKit's Apple Neural Engine implementation unlocks a much higher compute throughput than was available in \cite{redrafter}'s Apple Silicon GPU experiments, minimizing the \textit{verification overhead} to levels comparable to Nvidia GPUs.

However, we choose not to employ speculative decoding in our final WhisperKit implementation for Whisper Large v3 Turbo due to the \textit{drafter overhead}, the slowdown that is attributed to running the drafter model 4 times (beam length) in addition to running the original text decoder once.
For Whisper Large v3, this overhead is minimal due to the original model's high latency and the practical speedup ends up being 2.24x.
For Whisper Large v3 Turbo, this overhead is significant due to the original model's low latency and the practical speedup ends up being 1.25x.


\subsection{On-device Constraints}
There are two challenges to overcome before billion-scale Transformers can be ubiquitously deployed on edge devices: Energy and peak memory consumption.

\subsubsection{Energy}
Energy consumption directly impacts battery life and thermal sustainability of using an edge device and on-device inference must be extremely energy efficient in order to avoid disruptive patterns such as degraded battery life and too-hot-to-hold devices that throttle themselves to a halt in order to cool down.

In 2018, Apple introduced the Neural Engine (ANE) with the A12 chip for iPhone XS, an inference accelerator that optimized \textit{performance per watt} under 10 watts. In 2020, the Neural Engine was introduced to Mac and iPad with the M1 chip \cite{ane-transformers}. In 2023, all devices supported by iOS 17 had a Neural Engine. In 2025, iOS 17 is the oldest supported operating system for most commercial applications, making the Neural Engine ubiquitous for all Apple users.

In the Android market, Qualcomm, Mediatek, and others have driven a similar hardware transformation by introducing the Neural Processing Unit (NPU) in mobile chipsets such as Snapdragon 8 and Dimensity. In 2024, the Qualcomm Snapdragon X Elite chip was adopted by Microsoft in Copilot PCs, kickstarting the adoption of NPUs by Windows.

However, ANE and NPUs in general remain underutilized due to lack of public documentation and tooling. They have been primarily utilized by the device makers at the firmware level to power inference for operating system features such as computational photography \cite{hyperdetr} and speech recognition \cite{apple-asr-conformer}. In 2022, Apple published "Deploying Transformers on the Apple Neural Engine" and open-sourced \texttt{ane-transformers} \cite{ane-transformers} to demonstrate how a 0.1 billion parameter Transformer Encoder model could be deployed on the iPhone while being faster than server-side inference \cite{ane-transformers}. In 2024, Apple unveiled Apple Intelligence \cite{gunter2024appleintelligencefoundationlanguage} which is a 3-billion parameter Transformer language model running on the Neural Engine. The performance achieved by Apple Intelligence is not easily accessible to any third-party model due to lack of documentation. However, the necessary tooling is available with Core ML \cite{coreml}.

WhisperKit builds on the \texttt{ane-transformers} reference implementation of the Transformer architecture and applies additional optimizations to reach Apple Intelligence-level performance for a third-party Transformer model using the Core ML framework. In specific, \textit{Stateful Models} \cite{stateful} feature is leveraged to ensure that the key-value cache of the Whisper Text Decoder is read and updated in-place, persisting across forward passes. This leads to a 45\% latency reduction (8.4 ms \textrightarrow 4.6 ms on M3 ANE) for the Whisper Large v3 Turbo Text Decoder forward pass compared to our previous implementation that passed the key-value cache as input tensors to the model.

Most importantly, the energy consumption for one forward pass of the same model is reduced by 75\% (1.5W \textrightarrow 0.3W), mitigating the battery degradation and heating problem.



\begin{figure}[ht]
\begin{center}
    \begin{tikzpicture}[
      scale=0.8,
      transform shape,
      >=Stealth,
      every node/.style={font=\small},
      tensor/.style   ={draw, fill=gray!10,  minimum width=1.2cm, minimum height=1.2cm, align=center},
      inlier/.style   ={draw, fill=green!5, minimum width=1.2cm, minimum height=1.2cm, align=center},
      outlier/.style   ={draw, fill=red!5, minimum width=1.2cm, minimum height=1.2cm, align=center},
      sparse/.style   ={draw, fill=orange!15, minimum width=1.2cm, minimum height=1.2cm, align=center},
      quant/.style    ={draw, fill=blue!15,  rounded corners, minimum width=1.2cm, minimum height=1.2cm, align=center},
      stored/.style   ={draw, dashed, fill=none, rounded corners, minimum width=1.2cm, minimum height=1.2cm, align=center},
      plus/.style     ={circle, draw, fill=white, inner sep=0pt, minimum size=8pt},
      input/.style    ={draw, fill=yellow!15, minimum width=0.5cm, minimum height=0.5cm, align=center},
      output/.style   ={draw, fill=purple!15, minimum width=0.5cm, minimum height=0.5cm, align=center},
      op/.style      ={draw, fill=gray!50, minimum width=2.5cm, minimum height=0.5cm, align=center}
    ]
    
    \node[tensor] (w) {
      \begin{tikzpicture}[transform canvas={scale=0.4}, baseline=(mm.center)]
        \matrix (mm) [
          matrix of nodes,
          nodes={minimum size=0.4cm},
          row sep=0.05cm,
          column sep=0.05cm,
          matrix anchor=center
        ] {
          0 & 0 & 0 & 0 & 0 \\
          0 & 0 & 0 & 0 & 0 \\
          0 & 0 & 0 & 0 & 0 \\
          0 & 0 & 0 & 0 & 0 \\
          0 & 0 & 0 & 0 & 0 \\
        };
        
        \foreach \i in {1,...,5} {
          \foreach \j in {1,...,5} {
            \draw[green, thick, circle, fill=green!40] (mm-\i-\j) circle (0.15cm);
          }
        }
    
        \draw[red, thick, circle, fill=red!40] (mm-1-2) circle (0.15cm);
        \draw[red, thick, circle, fill=red!40] (mm-4-5) circle (0.15cm);
      \end{tikzpicture}
    };
    \node[below=0.05cm of w] (w_text) {$W$};

    \path (w.east) ++(1.6,  1) node[inlier] (in)  {
      \begin{tikzpicture}[transform canvas={scale=0.4}, baseline=(mm_in.center)]
        \matrix (mm_in) [
          matrix of nodes,
          nodes={minimum size=0.4cm},
          row sep=0.05cm,
          column sep=0.05cm,
          matrix anchor=center
        ] {
          0 & 0 & 0 & 0 & 0 \\
          0 & 0 & 0 & 0 & 0 \\
          0 & 0 & 0 & 0 & 0 \\
          0 & 0 & 0 & 0 & 0 \\
          0 & 0 & 0 & 0 & 0 \\
        };
        
        \foreach \i in {1,...,5} {
          \foreach \j in {1,...,5} {
            \ifnum\i=1
              \ifnum\j=2
              \else
                \draw[green, thick, circle, fill=green!40] (mm_in-\i-\j) circle (0.15cm);
              \fi
            \else
              \ifnum\i=4
                \ifnum\j=5
                \else
                  \draw[green, thick, circle, fill=green!40] (mm_in-\i-\j) circle (0.15cm);
                \fi
              \else
                \draw[green, thick, circle, fill=green!40] (mm_in-\i-\j) circle (0.15cm);
              \fi
            \fi
          }
        }
      \end{tikzpicture}
    };
    \node[below=0.05cm of in] (in_text) {$W_{\text{inlier}}$};
    \path (w.east) ++(1.6, -1) node[outlier] (out) {
      \begin{tikzpicture}[transform canvas={scale=0.4}, baseline=(mm_in.center)]
        \matrix (mm_in) [
          matrix of nodes,
          nodes={minimum size=0.4cm},
          row sep=0.05cm,
          column sep=0.05cm,
          matrix anchor=center
        ] {
          0 & 0 & 0 & 0 & 0 \\
          0 & 0 & 0 & 0 & 0 \\
          0 & 0 & 0 & 0 & 0 \\
          0 & 0 & 0 & 0 & 0 \\
          0 & 0 & 0 & 0 & 0 \\
        };
        
        \foreach \i in {1,...,5} {
          \foreach \j in {1,...,5} {
            \ifnum\i=1
              \ifnum\j=2
                \draw[red, thick, circle, fill=red!40] (mm_in-\i-\j) circle (0.15cm);
              \else
              \fi
            \else
              \ifnum\i=4
                \ifnum\j=5
                  \draw[red, thick, circle, fill=red!40] (mm_in-\i-\j) circle (0.15cm);
                \else
                \fi
              \else
              \fi
            \fi
          }
        }
      \end{tikzpicture}
    };
    \node[below=0.05cm of out] (out_text) {$W_{\text{outlier}}$};

    \draw[->] (w.east) to[out= 10,in=180] (in.west);
    \draw[->] (w.east) to[out=-10,in=180] (out.west);

    \node[quant, right=1.8cm of in] (q) {$\mathcal{Q}(W_{\text{inlier}})$};
    \draw[->] (in.east) -- node[above] {palettize}(q.west);

    \node[tensor, right=1.8cm of out] (s) {$\mathcal{S}(W_{\text{outlier}})$};
    \draw[->] (out.east) -- node[above] {sparsify}(s.west);

    \node[stored, fit=(q) (s), inner sep=7pt] (stored) {};
    \node[below=0.05cm of stored] (stored_text) {$W_{\text{OD-MBP}}$};

  \end{tikzpicture}
\caption{
  \textbf{Compile time with OD-MBP.} 
  The original weight tensor $\mathbf{W}$ is decomposed into a dense inlier block 
  $\mathbf{W}_{\text{inlier}}$ (green) and a sparse outlier block 
  $\mathbf{W}_{\text{outlier}}$ (red). Outliers are defined as weight values that are more than 3 standard deviations away from the mean.
  Inliers are \emph{palettized} and stored as a low-bit lookup table (LUT) as $Q(\mathbf{W}_{\text{inlier}})$, 
  while outliers are kept in float16 precision and stored in a bit-packed sparse representation as $S(\mathbf{W}_{\text{outlier}})$.
}

\label{OD-MBP-compile-time}
\end{center}
\vskip -0.25in
\end{figure}
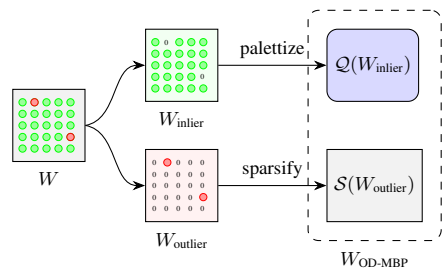

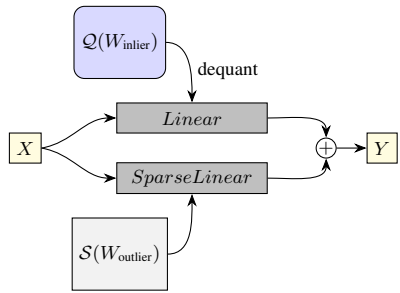
\begin{figure}[ht]
  \begin{center}
      \begin{tikzpicture}[
        scale=0.8,
        transform shape,
        >=Stealth,
        every node/.style={font=\small},
        tensor/.style   ={draw, fill=gray!10,  minimum width=1.2cm, minimum height=1.2cm, align=center},
        inlier/.style   ={draw, fill=green!5, minimum width=1.2cm, minimum height=1.2cm, align=center},
        outlier/.style   ={draw, fill=red!5, minimum width=1.2cm, minimum height=1.2cm, align=center},
        sparse/.style   ={draw, fill=orange!15, minimum width=1.2cm, minimum height=1.2cm, align=center},
        quant/.style    ={draw, fill=blue!15,  rounded corners, minimum width=1.2cm, minimum height=1.2cm, align=center},
        stored/.style   ={draw, dashed, fill=none, rounded corners, minimum width=1.2cm, minimum height=1.2cm, align=center},
        plus/.style     ={circle, draw, fill=white, inner sep=0pt, minimum size=8pt},
        input/.style    ={draw, fill=yellow!15, minimum width=0.5cm, minimum height=0.5cm, align=center},
        output/.style   ={draw, fill=purple!15, minimum width=0.5cm, minimum height=0.5cm, align=center},
        op/.style      ={draw, fill=gray!50, minimum width=2.5cm, minimum height=0.5cm, align=center}
      ]
  
      \node[input] (x) {$X$};
      \path (x.east) ++(2.5,  0.5) node[op] (linear) {$Linear$};
      \path (x.east) ++(2.5, -0.5) node[op] (sparse) {$SparseLinear$};
      \draw[->] (x.east) to[out= 10,in=180] (linear.west);
      \draw[->] (x.east) to[out=-10,in=180] (sparse.west);
  
      \node[quant] (q_inference) at ($(linear.north)+(-1.2cm,1.0cm)$) {$\mathcal{Q}(W_{\text{inlier}})$};
      \node[tensor] (s_inference) at ($(sparse.south)+(-1.2cm,-1.0cm)$) {$\mathcal{S}(W_{\text{outlier}})$};
      \draw[->] (q_inference.east) to[out=0,in=90] node[right, pos=0.6] {dequant} (linear.north);
      \draw[->] (s_inference.east) to[out=0,in=-90] (sparse.south);
  
      \node[plus] (add) at ($(x)+(5,0)$) {$+$};
      \draw[->] (linear.east) to[out=0,in=90] (add.north);
      \draw[->] (sparse.east) to[out=0,in=-90] (add.south);
  
      \node[input, right=0.5cm of add] (y) {$Y$};
      \draw[->] (add.east) -- (y.west);
  
    \end{tikzpicture}
  \caption{
    \textbf{Inference with OD-MBP.}
    The float16 input activation \(X\) is dispatched to two parallel operators:  
    \emph{(i)~Dense path.} A \texttt{Linear} layer retrieves the low-bit palettized inlier weights \(Q(W_{\text{inlier}})\), dequantizes them on the fly, and performs a dense matrix–vector product in float16 precision.  
    \emph{(ii)~Sparse path.} A \texttt{SparseLinear} layer retrieves the float16 precision sparse outlier weights \(S(W_{\text{outlier}})\) and performs a sparse matrix–vector product in float16 precision.
      }
  \label{OD-MBP-runtime}
  \end{center}
  \vskip -0.33in
  \end{figure}


  \subsubsection{Memory}
  The model size introduces different bottlenecks throughout the lifecycle of on-device deployment. First, the model file must be distributed to the end-user device over the air (OTA) \textit{after} the app is installed to keep the app size small and decouple model updates from software updates. The only increase in the app size must be tied to the on-device inference SDK which is generally less than 5 MB. Once distributed, the model files take up storage space even when not being used, so models larger than 2 GB are not end-user friendly.
   
  Peak memory consumption during active use is dominated by the model weights and models optimized for on-device deployment must require less than 2 GB of RAM in order to retain virtually universal device support. This size roughly corresponds to a 1-billion parameter model in float16 precision or 4-billion parameter model in 4-bit precision after weight compression. Whisper Large v3 Turbo requires 1.6 GB in float16 precision. Naive compression of Whisper models lead to increased Word Error Rate (WER) along with hallucination and repetitive patterns in the predicted transcription.
  
  WhisperKit leverages a new compression technique called \textit{Outlier-Decomposed Mixed-Bit Palletization} (OD-MBP) to retain WER within 1\% of the original uncompressed model while staying below 1 GB in size. \textit{Mixed-Bit Palettization} (MBP) was first proposed in \cite{mixedbitpalettization} to compress Stable Diffusion XL with the natively accelerated compression format called \textit{palettization} \cite{coreml-compression}. Subsequently, MBP was adopted and improved upon by others such as \cite{bitsfusion} to achieve low-bit high-accuracy weight compression for edge deployment.
  
  Outlier Decomposition (OD) was first proposed in \cite{llmint8} where the forward pass of a linear layer is decomposed into a low-bit precision \textit{inlier} branch and a float16 precision \textit{outlier} branch to minimize the errors caused by low-bit compression. Outliers are determined based on the channel-wise magnitude statistics. By definition, they represent less than 1\% of the data. However, this technique does not accelerate inference because the latency gains from low-bit compressed weights requiring less memory bandwidth are offset by the additional compute from the float16 branch. Furthermore, this technique is restricted to feature-wise structured decomposition and does not admit unstructured outlier decomposition.

  OD-MBP builds on top of OD and MBP by decomposing the weight tensors as shown in Figure \ref{OD-MBP-compile-time} such that the low-bit precision inlier branch is implemented using the natively-accelerated palettization format and the float16 precision outlier branch is implemented using the natively-accelerated sparse weight format \cite{coreml-compression} that is capable of accelerating almost fully unstructured sparsity \cite{coreml-compression}. Figure \ref{OD-MBP-runtime} shows changes to the forward pass of a \texttt{Linear} layer when its weights are compressed with OD-MBP.

  When applied to Whisper Large v3 Turbo, OD-MBP preserves the WER of the original model to within 1\% across various short- and long-form datasets as shown in Table \cref{compressed-causal-wer}.

\begin{table}[ht]
  \begin{center}
    {\scalebox{0.45}{
    \begin{tabular}{lccccc}
      \toprule
      Compression State & Mask & Size [GB] & \makecell{librispeech-test.clean \\ (WER)} & \makecell{earnings22-12hours \\ (WER)} & \makecell{CommonVoice17-en \\ (WER)} \\
      \midrule
      FP16 & original & 1.6 & 1.93 & 11.55 & 12.13 \\
      OD-MBP & original & 0.6 & 1.96 (+0.03) & 12.35 (+0.80) & 13.03 (+0.90) \\
      FP16 & d750 & 1.6 & 2.25 (+0.32) & 12.85 (+1.30) &  12.87 (+0.74) \\
      OD-MBP & d750 & 0.6 & 2.30 (+0.37) & 12.72 (+1.17) &  14.21 (+2.08) \\
      \bottomrule
      \end{tabular}
    }}
  
  \caption{\textbf{Impact of OD-MBP on WER}. OD-MBP, even when combined with the \texttt{d750} optimization, retains WER within 1\% of the original uncompressed model while compressing Whisper Large v3 Turbo from 1.6 GB to 0.6 GB}
  \label{compressed-causal-wer}
  \end{center}
  \vskip -0.3in
  \end{table}

\section{Results}

In this section, WhisperKit is benchmarked to diverse and competitive cloud-based inference APIs for real-time transcription.OpenAI \texttt{gpt-4o-transcribe} \cite{gpt-4o-transcribe} sets the frontier model baseline. This particular version of \texttt{gpt-4o} is fine-tuned for transcription and sets an even stronger baseline than the base \texttt{gpt-4o}. Deepgram \texttt{nova-3} \cite{deepgram-nova-3} \cite{deepgram-nova-3} sets the baseline for proprietary ASR-focused models. Fireworks \texttt{large-v3-turbo} \cite{fireworks-large-v3-turbo} sets the baseline for ASR-focused models using the same open-source model as WhisperKit (Whisper Large v3 Turbo), serving it on the cloud instead of on device.
\subsection{Latency}

Real-time transcription latency \footnote{SDBench is used for measurements. Code available at: \url{https://github.com/argmaxinc/SDBench}} is measured as the difference between the \textit{audio cursor} and the \textit{transcript cursor} following Deepgram's methodology \cite{deepgram-latency} with one improvement: In order to remove the impact of predicted word-level timestamps when setting the \textit{transcript cursor}, we run our benchmarks on the TIMIT dataset \cite{timit} and adopt its ground-truth word-level timestamps for this purpose.

\begin{figure}[ht]
\vskip -0.08in
\begin{center}
\includegraphics[width=\linewidth]{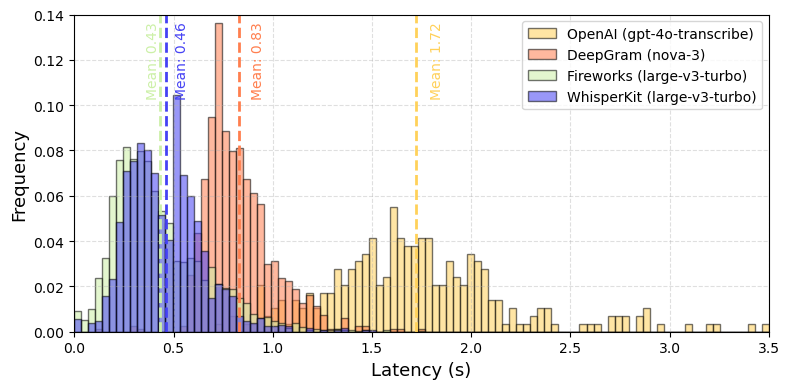}
\caption{\textbf{Per-word Latency Histogram for Hypothesis Text.} Hypothesis text is also referred to as \textit{Interim Result} in Deepgram's documentation \cite{deepgram-latency}. Vertical dashed lines indicate the mean latency for each system.}
\label{fig:per-word-latency-hypothesis}
\end{center}
\vskip -0.2in
\end{figure}

This improvement was also necessary because Fireworks \cite{fireworks-large-v3-turbo} and OpenAI \cite{gpt-4o-transcribe} APIs do not return predicted word-level timestamps. We verified that our latency results in Figure \ref{fig:per-word-latency-hypothesis} for Deepgram and Fireworks approximately match their offically reported API latency in \cite{300m-fireworks} and \cite{deepgram-latency} despite our metric improvement.

\begin{figure}[ht]
  \begin{center}
  \includegraphics[width=\linewidth]{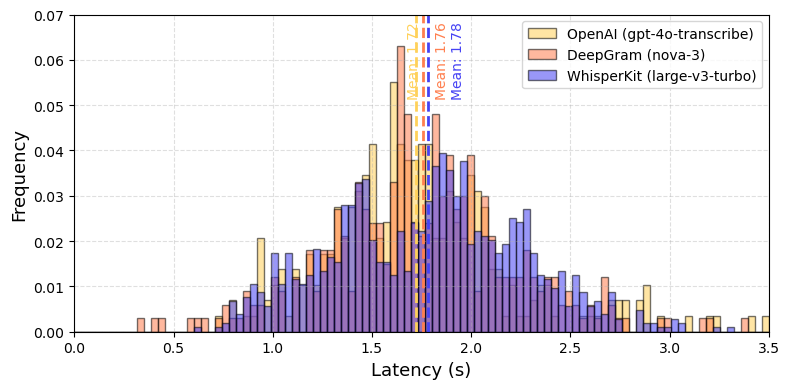}
  \caption{\textbf{Per-word Latency Histogram for Confirmed Text.} When \textit{hypothesis text} results are not considered, all benchmarked systems achieve a similarly high latency of ~1.7 seconds. Fireworks API does not support partial resul confirmations for any result and is excluded from this plot.}
  \label{fig:per-word-latency-confirmed}
  \end{center}
 \vskip -0.14in
  \end{figure}

Figure \ref{fig:per-word-latency-hypothesis} shows the per-word latency for the \textit{hypothesis text} stream. WhisperKit and Fireworks are the fastest, achieving a roughly equal mean latency of ~0.45 seconds. Deepgram achieves 0.83 seconds and becomes the third fastest system. Note that OpenAI does not support \textit{hypothesis text} streams so the results represent the latency of the \textit{confirmed text} stream, which is expectedly much slower than all other systems. Figure \ref{fig:per-word-latency-confirmed} shows the per-word latency for the \textit{confirmed text} stream. All systems achieve a similar latency of ~1.7 seconds. Fireworks is not shown on this figure because their API does not confirm the transcription in contiguous segments.

\subsection{Accuracy}

Evaluating the accuracy of real-time transcription systems is challenging due to the presence of retrospective corrections throughout real-time streaming inference. For comprehensive evaluation with and without retrospective corrections, we designed the following evaluation framework:

\begin{itemize}
  \item \textit{Confirmed text} accuracy is evaluated once the streaming session (per file) is complete. We leverage the standard Word Error Rate (WER) metric for evaluating this text stream (lower is better). Systems are allowed to correct past mistakes while a result is still marked as \textit{hypothesis text}. Systems are not allowed to retrospectively change {confirmed text}. 
  \item \textit{Hypothesis text} accuracy is measured by the number of corrections made by the system before confirmation. If number of corrections is zero, then \textit{hypothesis text} accuracy achieves its upper bound and matches the accuracy of \textit{confirmed text}. If the number of corrections is too high, then the reduced latency afforded by \textit{hypothesis text} is not useful and the user experience is degraded due to unstable results.
\end{itemize}

\begin{figure}[ht]
\vskip - 0.15in
\begin{center}
\includegraphics[width=\linewidth]{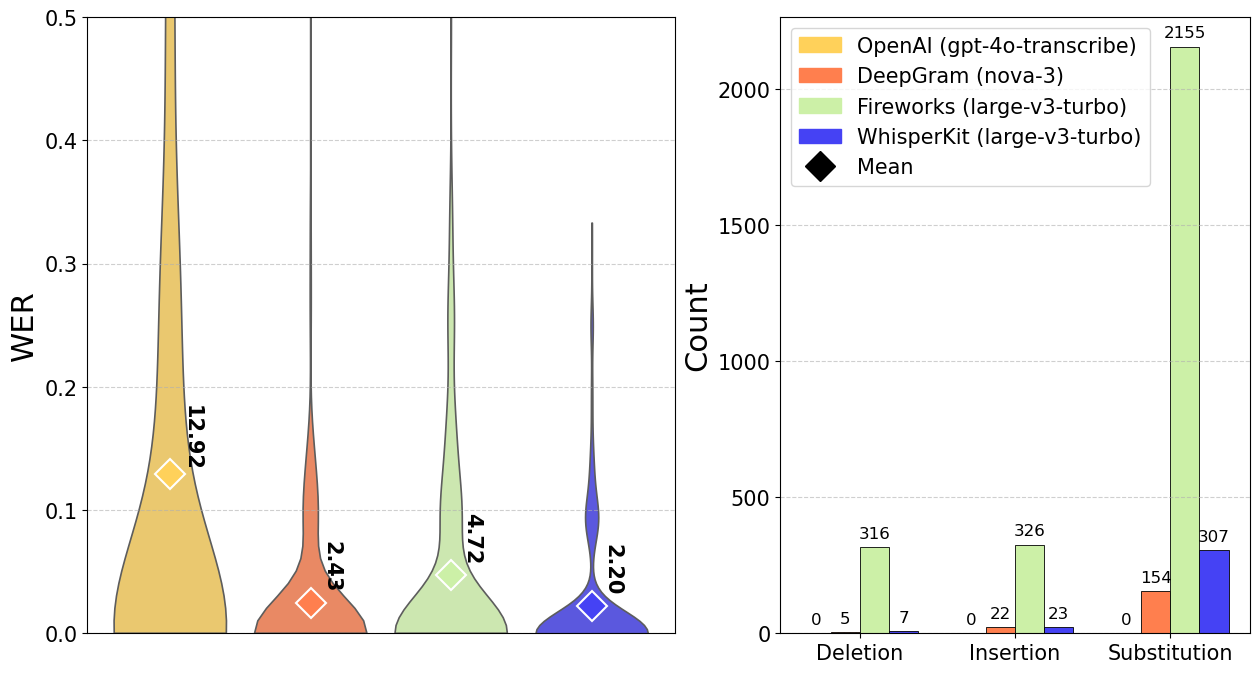}
\caption{\textbf{Streaming Real-time Transcription Accuracy.} On the left, \textit{confirmed text} accuracy is shown as WER plots (lower is better). On the right, \textit{hypothesis text} accuracy is shown as the total number of corrections (lower is better) stratified across deletions, substitutions, and insertions.}
\label{fig:streaming-wer}
\end{center}
\vskip -0.2in
\end{figure}

Figure \ref{fig:streaming-wer} shows that OpenAI's \textit{confirmed text} accuracy is lowest (highest WER). WhisperKit and Deepgram achieve the highest accuracy (lowest WER) of 2\% WER while Fireworks trails behind at 4.72\% WER.  Notably, OpenAI demonstrates zero corrections because the API does not support \textit{hypothesis text}. On the other hand, Fireworks issues an order of magnitude more corrections than WhisperKit and Deepgram. Despite tying with WhisperKit as the lowest latency system in Figure \ref{fig:per-word-latency-hypothesis}, Fireworks' extremely high number of corrections makes it undesirable for real-time applications. High number of corrections may be associated with small and less accurate models predicting ahead of large and more accurate models. WhisperKit latency is measured on a MacBook Pro with M3 Max chip on the Neural Engine.

\label{submission}

\bibliography{example_paper}
\bibliographystyle{icml2025}




\end{document}